\documentclass[epsfig,aps,pra,twocolumn]{revtex4}

\usepackage{amssymb}
\usepackage{amsmath}
\usepackage{graphicx}
\usepackage{dcolumn}
\usepackage{bm}

\begin{document}
\newcommand{\eexp}[1]{\mbox{e}^{#1}}
\title{Particle Motion in Rapidly Oscillating Potentials: The Role of the Potential's Initial Phase}
\author{A. Ridinger and N. Davidson.}
\affiliation{Department of Physics of Complex Systems, Weizmann
Institute of Science, Rehovot 76100, Israel}
\date{\today}

\begin{abstract}
Rapidly oscillating potentials with a vanishing time average have
been used for a long time to trap charged particles in source-free
regions. It has been argued that the motion of a particle inside
such a potential can be approximately described by a time
independent effective potential, which does not depend upon the
initial phase of the oscillating potential. However, here we show that 
the motion of a particle and its trapping condition significantly depend 
upon this initial phase for arbitrarily high frequencies of the potential's oscillation. We explain 
this novel phenomenon by showing that the motion of a particle 
is determined by the effective potential stated in the 
literature only if its initial conditions are transformed according to a transformation which we show to significantly depend on the potential's initial phase for arbitrarily high frequencies. We confirm our theoretical findings
by numerical simulations. Further, we demonstrate that the found phenomenon
offers new ways to manipulate the dynamics of particles which are trapped 
by rapidly oscillating potentials. Finally, we propose a simple experiment 
to verify the theoretical findings of this work.\\
\end{abstract}
 \maketitle

\section{Introduction}
The theoretical study of the dynamics of a particle moving in
a rapidly oscillating potential has been an active field of
research for the past 50 years
\cite{Kap51,LanLif76,GapMil58,Deh67,PerRic82,NadOks86,Com86,CooSha85,GroRak88,Fis03}.
In 1951, Kapitza showed that rapidly oscillating potentials
can have a stabilizing effect on dynamical systems \cite{Kap51},
which he demonstrated utilizing an inverted pendulum whose
suspension point was forced to oscillate vertically. For high
enough frequencies and large enough amplitudes of this oscillation
the upwards position of the pendulum becomes stable. Kapitza
showed that the pendulum feels a time independent \textit{effective} potential
that has a local minimum at the pendulum's upwards position and
thus imparts stability upon it. This phenomenon has since been
termed ``dynamic stabilization''. In 1958, Paul used the
principle of dynamic stabilization to trap charged ions by
alternating electric fields \cite{Pau90}. Ions trapped within a
Paul trap have and continue to be used in the development of
frequency standards, other fundamental measures \cite{BlaGil92} and quantum information \cite{CirZol95}.\\
\indent In order to grasp all the possible applications that are
offered by atoms which are trapped by rapidly oscillating
potentials, a descriptive theory of their dynamics is
indispensable. For the motion of ions in a Paul trap such a
theory is given by Mathieu's theory, which provides the exact
solution of the ion's classical equations of motion in terms of
the tabulated Mathieu functions \cite{Pau90,LeiBla03,MajGhe04}.
However, this theory is restricted to potentials of harmonic shape
which are oscillated sinusoidally only. Besides this, Mathieu's
theory does not raise intuition regarding the atom's dynamics
hindering the discovery of new phenomena. For these purposes,
Kapitza's theory of the effective potential has thus far been
suitable, sufficiently explaining the stabilization of the
inverted pendulum \cite{Kap51}. This theory has since been
generalized to arbitrary rapidly oscillating potentials
\cite{LanLif76,Fis03} and extended to quantum mechanics
\cite{CooSha85, GroRak88, Fis03}.\\
\indent In this paper, we describe a different phenomenon, which illustrates that the motion of a particle trapped by a rapidly oscillating potential with a vanishing time average significantly depends on the potential's initial phase for arbitrarily high frequencies of its oscillation. We discover that this phenomenon cannot be explained by the effective potential theory which we find to be incomplete. Therefore, we derive the necessary completion in order to describe this phenomenon.\\ 
\indent Kapitza's effective potential theory is based on a separation of the particle's motion into a slow and a fast changing part. For high frequencies of the potential's oscillation this separation can be performed in such a way that the slow part is governed by a time independent potential while the fast part has a vanishing time average and a negligible amplitude. In Kapitza's theory it is concluded that the particle's real motion is approximately described by its slow part, such that the real motion can be obtained by replacing its time dependent equation of motion by the time independent equation of motion of the slow part. However, doing so requires replacing the particle's initial conditions by the corresponding initial conditions of the slow part. If these two sets of initial conditions significantly differ from each other, the replacement of the initial conditions is a crucial point in the determination of the real motion. In the context of adiabatic elimination a significant difference between the initial conditions of the full problem and the model problem is called an ``initial slip'' \cite{GeiTit83,HaaLew83,CoxRob95}.\\ 
\indent We show in our problem that there is an initial slip even for arbitrarily high frequencies of the potential's oscillation. As a result, the approximate determination of the particle's real motion must be performed in two independent steps: Firstly the real initial conditions have to be transformed and then the time independent equation of motion of the slow part must be solved for these transformed initial conditions. In Kapitza's theory the transformation of the initial conditions is omitted. We explicitly derive this transformation and show that it significantly depends on the oscillating potential's initial phase for arbitrarily high frequencies. This explains the found phenomenon that the particle's motion depends on the initial phase.\\
\indent The phenomenon is clearly manifested in the particle's stability region (i.e. the region in phase space that contains all initial conditions of the particle leading to subsequent trapping by the oscillating potential) for which we derive an analytical formula. Furthermore, we show that the phenomenon offers the possibility to manipulate particles, that are trapped by rapidly oscillating potentials, in a simple and controlled manner. For example, modification of the oscillating potential's phase at some point in time by a discrete value (inducing a phase hop) results in a change of the trapped particle's average energy. We explicitly calculate this change in energy and show that it is significant for arbitrarily high frequencies of the potential's oscillation. For the special case of a sinusoidally oscillated arbitrary potential we also show that -- in the framework of classical physics -- it is significant even for arbitrarily low energies of the trapped particle.\\
\indent We confirm all theoretical findings in this paper by numerical simulations. For this purpose, we consider two simple examples: A particle moving inside an oscillating Gaussian-shaped potential and inside an oscillating parabola-shaped potential. We also propose a simple experiment to validate our findings.\\ 
\indent Throughout the paper we use one-dimensional formalism, but
the results presented here can be directly applied to two and
three dimensions in cases where the motion is separable
\cite{LanLif76}. For high dimensional oscillating potentials where
the particle's motion in not separable and in particular chaotic,
new phenomena may occur that are not discussed here
\cite{FriKap01}.

\section{Particle Motion in a Rapidly Oscillating Potential}
\indent In this section, we derive an approximate theory for the motion of a particle that is moving inside a rapidly oscillating potential. This is done in two steps: Firstly, we formulate a model problem with a time independent ``effective'' equation of motion and then we show that this model problem possesses initial slip and derive the necessary transformation of the real problem's initial conditions to the corresponding initial conditions of the model. With the obtained theory we then verify the stated phenomenon of the dependence of the particle's motion on the initial phase of the oscillating potential. Finally, we describe a clear manifestation of this phenomenon and its main consequences.\\
\indent Newton's equation for a particle with mass $m$ that is
moving inside a one-dimensional rapidly oscillating potential
$V(x,t)$ can be written as
\begin{eqnarray}\label{realeom}
   m\ddot{x}=-V'(x,t)=-V'_0(x)-V_1'(x)f(\omega t\!+\!\varphi),
\end{eqnarray}
where $V_0(x)$ describes the time average part of $V(x,t)$ and $V_1(x)$ its oscillating part and $f(\omega t)$ its time modulation function
which has period $2\pi$, unity amplitude and vanishing time average over one period. The frequency of the potential's oscillation is denoted by
$\omega$. $\varphi$ is  the initial phase of the oscillating potential $V(x,t)$. Derivatives with respect to coordinates are denoted by primes
and with respect to time by dots. The average over one period of oscillation (period-average) of a $2\pi$-periodic function $g(\tau)$ is denoted by a
bar: $\overline{g}\equiv\,$\mbox{$\frac{1}{2\pi}\int_0^{2\pi}g(\tau)\,d\tau$}. If the frequency $\omega$ is very large compared to the other time scales of the system, the time dependent equation of
motion (\ref{realeom}) can be replaced by a time independent \textit{effective} equation of motion that yields the period-averaged motion of the
particle \cite{Fis03}. This effective equation of motion can be obtained by separating the motion $x(t)$ of the particle into a sum of a slow part
$X(t)$, which depends on the slow time $t$, and a fast part $\xi(\tau)$, which depends on the fast time $\tau\equiv\omega t$ \cite{Kap51}:
\begin{eqnarray}\label{Separation}
x(t)=X(t)+\xi(\tau).
\end{eqnarray}
$X(t)$ and $\xi(\tau)$ will be referred to as the slow and the
fast motion, respectively. For large frequencies $\omega$, the
amplitude of the fast motion $\xi$ can be assumed to be small,
since due to its inertia, the particle does not have the time to
react to the force which is induced by the oscillating potential,
before this force changes sign. According to \cite{Fis03}, $\xi$
can for large frequencies thus be expanded in powers of the inverse frequency:
$\xi(\tau)=\sum_{i=1}^{\infty}\frac{1}{\omega^i}\xi_i(\tau)$.
Requiring that on the fast time scale $\tau$, the fast motion
$\xi$ is periodic with a vanishing period-average, the particle's
period-averaged motion \mbox{$\overline{x(t)}$} is given by the
slow motion, $\overline{x(t)}=X(t)$. Further requiring that the
definition of $\xi$ leads to a time independent equation of motion
for the slow motion $X(t)$, uniquely defines $\xi$ and yields the
effective equation of motion for $X(t)$. This calculation is
presented in Ref$.$ \cite{Fis03}. Its results in the leading
order in $\omega^{-1}$ are the explicit expression of $\xi$,
\begin{eqnarray}\label{xi}
  \xi(\tau,X,\varphi)= -\frac{1}{m\omega^2}V_1'(X)\int^{(2)\tau}[f(\tau\!+\!\varphi)],
\end{eqnarray}
and the effective equation of motion
\begin{eqnarray}\label{effeomfishman}
m\ddot{X}=-V_{\textrm{\scriptsize eff}}'(X)
\end{eqnarray}
with the effective potential
\begin{eqnarray}\label{Effectivepotential}
V_{\textrm{\scriptsize eff}}(X)=V_0(X)+\frac{1}{2m\omega^2}V_1'(X)^2\overline{\left(\int^\tau [f(\tau)]\right)^2}.
\end{eqnarray}
The standardized integral $\int^\tau [f(\tau)]$ is the anti-derivative of $f(\tau)$ with a vanishing
period-average \cite{SanVer85}. For $2\pi$-periodic
functions $f(\tau)$ that have a vanishing period-average and whose
Fourier expansion is given by $f(\tau)=\sum_{n\neq0}f_n
e^{in\tau}$, it is defined as $\int^\tau [f(\tau)]\equiv
\sum_{n\neq0}\frac{1}{in}f_ne^{in\tau}$. It can be applied repeatedly. Its multiple
application ($j$ times) is denoted by $\int^{(j)\tau} [f(\tau)]$
and its evaluation at the point $\tau_0$ by $[\int^\tau
[\cdots]]_{\tau_0}$.\\
\indent The effective equation of motion (\ref{effeomfishman}) defines a simplified model of the original problem. We now show that it possesses initial slip. For that matter, we compare both the position $x(t)$ with its
period-average $X(t)$ and the velocity $\dot{x}(t)$ with its
period-average $\dot{X}(t)$ determined by the original and the model problem, respectively. (For a numerical example see Sec$.$
III, Fig$.$ \ref{Trajectories}.) We restrict ourselves to the
physically interesting case of a particle that is trapped by a
rapidly oscillating potential with a vanishing time average (i.e.
$V_0\equiv0$). The position $X(t)$ and the velocity $\dot{X}(t)$
of the period-averaged motion are determined by equation
(\ref{effeomfishman}). Thus, for the particle to be trapped, the
effective potential (\ref{Effectivepotential}) must have a local
minimum. The spatial width of this local minimum determines the
spatial confinement of the particle, and is, according to Eq$.$
(\ref{Effectivepotential}), independent of $\omega$. Thus, the
position $X(t)$ of the trapped particle's slow motion is of the
order $\omega^{0}$. According to (\ref{xi}), the position $\xi(t)$
of its fast motion is of the order $\omega^{-2}$ and therefore in
the limit of large $\omega$ negligible compared to the position
$X(t)$ of its slow motion. This means that, for
large $\omega$, one can approximate $x(t)=X(t)+\xi(t)\approx X(t)$ (see Fig$.$ \ref{Trajectories}(a)).\\
\indent The depth of the effective potential's local minimum determines which kinetic energies the trapped particle can have. It is, according to Eq$.$
(\ref{Effectivepotential}), of the order $\omega^{-2}$. Thus, the velocity $\dot{X}(t)$ of the particle's slow motion is of the order
$\omega^{-1}$. The velocity $\dot{\xi}(t)$ of its fast motion is given by the derivative of Eq$.$ (\ref{xi}) with respect to time,
\begin{eqnarray}\label{xidot}
\dot{\xi}(t,\varphi)&=&-\frac{1}{m\omega}V_1'(x(t))\left[\int^{\tau}[f(\tau\!+\!\varphi)]\right]_{\tau=\omega
t},
\end{eqnarray}
which is a term of the order $\omega^{-1}$ and thus of the same order as $\dot{X}(t)$. Therefore, the velocity $\dot{\xi}(t)$ of the trapped
particle's fast motion is for large $\omega$ not negligible compared to the velocity $\dot{X}(t)$ of its slow motion. As a result, even for
large $\omega$, it is $\dot{x}(t)=\dot{X}(t)+\dot{\xi}(t)\not\approx\dot{X}(t)$ (see Fig$.$ \ref{Trajectories}(b)), such that in general it is $\dot{x}(0)\not\approx\dot{X}(0)$. This shows that there is an initial slip.\\
\indent This observation explains that in order to determine the particle's period-averaged motion with the effective equation of motion (\ref{effeomfishman}) for given initial conditions $x(0)$ and $\dot{x}(0)$ the initial condition transformation
\begin{eqnarray}\label{Trafo}
X(0)&=& x(0),\nonumber\\
\dot{X}(0)&=& \dot{x}(0)-\dot{\xi}(0),
\end{eqnarray}
with
\begin{eqnarray}\label{xi0dododo}
\dot{\xi}(0)&=&-\frac{1}{m\omega}V_1'(x(0))\left[\int^{\tau}[f(\tau\!+\!\varphi)]\right]_{\tau=0}
\end{eqnarray}
has to be performed, whereas the contribution of $\dot{\xi}(0)$ in Eq$.$ (\ref{Trafo}) cannot
be neglected.

\indent Expression (\ref{xi0dododo}) depends on the term \mbox{$\left[\int^{\tau}[f(\tau\!+\!\varphi)]\right]_{\tau=0}$}, which is a function of
the initial phase $\varphi$ of the oscillating potential $V(x,t)$. The initial condition transformation (\ref{Trafo}) therefore depends on
$\varphi$, which implies that the resulting period-averaged motion $X(t)$ of the particle also depends on $\varphi$ (For a numerical example, see Sec$.$ III, Fig$.$ \ref{Trajectoriesdifferentphi}). Further inspection of
expression (\ref{xi0dododo}) shows, that it also depends on the initial position $x(0)$ of the particle. This implies that the degree of
dependence of the particle's motion on the initial phase is determined by the particle's initial position. In the special case that the term
\mbox{$\left[\int^{\tau}[f(\tau\!+\!\varphi)]\right]_{\tau=0}$} vanishes, the initial conditions of the slow motion, $X(0),\dot{X}(0)$, equal
the real initial conditions, $x(0),\dot{x}(0)$, and there is no initial slip. In this case, the motion of the particle is determined by the time independent effective potential (\ref{Effectivepotential}) alone. Since $\int^{\tau}[f(\tau)]$ is
$2\pi$-periodic and has vanishing average, there always exist two phases $\varphi_1,\varphi_2\in[0,2\pi]$ for which
\mbox{$\left[\int^{\tau}[f(\tau\!+\!\varphi_1)]\right]_{\tau=0}$} and \mbox{$\left[\int^{\tau}[f(\tau\!+\!\varphi_2)]\right]_{\tau=0}$} vanish.
Thus, for every given rapidly oscillating potential, there exist two phases for which the motion of any particle inside this potential is
determined only by its time-independent effective
potential.\\
\indent The significance of the highlighted dependence of a
particle's motion on the initial phase is manifested in the
particle's stability region, which illustrates the particle's trapping condition. The stability region is the smallest region in phase space that contains all initial conditions of the particle leading to subsequent trapping by the oscillating potential. Since the effective potential (\ref{Effectivepotential}) in equation (\ref{effeomfishman}) is time independent, the principle of energy conservation can be invoked for its calculation. For a trapping
effective potential we define $X=0$ as the position of the bottom
of the trap and $X=L$ as the position of the trap boundary. Thus,
the initial conditions $X(0),\dot{X}(0)$ belong to the stability
region if they obey the inequality
\begin{eqnarray}\label{Trapcond}
   \frac{1}{2}m\dot{X}(0)^2+V_{\textrm{\scriptsize eff}}(X(0))\leq V_{\textrm{\scriptsize eff}}(L).
\end{eqnarray}
Substituting Eq$.$ (\ref{Trafo}), solving for $\dot{x}(0)$, and
neglecting terms of higher order than $\omega^{-1}$ yields
\begin{eqnarray}\label{Thesis}
\!\!\!\!\!\!\!\!\dot{x}(0)\!\! &\lessgtr&\!\!
-\frac{1}{m\omega}V_1'(x(0))\left[\int^{\tau}[f(\tau\!+\!\varphi)]\right]_{\tau=0}\nonumber\\
&&\!\pm\frac{1}{m\omega}\sqrt{\overline{\left(\int^\tau
[f(\tau)]\right)^2}}\sqrt{V_1'(L)^2-V_1'(x(0))^2},\ \ \ \
\end{eqnarray}
which defines the stability region for large $\omega$. The right-hand side of Eq$.$ (\ref{Thesis}) consists of a phase dependent and a
phase independent term which are both of the same order, such that
the stability region significantly depends on the phase $\varphi$
for arbitrarily
large $\omega$. In Sec$.$  III, we explicitly calculate the stability region for different $\varphi$ for two numerical examples.\\
\indent In the preceding paragraphs we showed that, due to the dependence of the fast motion's initial velocity $\dot{\xi}(0)$ on the initial phase $\varphi$ and due to the significant coupling between the particle's slow motion $X(t)$ and $\dot{\xi}(0)$, the particle's slow motion $X(t)$ is significantly coupled to $\varphi$. $\varphi$ can therefore be used to manipulate the particle's motion. For
example, if  $\varphi$ is modified instantaneously by a discrete value $\Delta\varphi$ (a phase hop), the position $X(t)$ of the particle's slow
motion will, on the time scale of the slow motion, change in a non-smooth manner and the velocity $\dot{X}(t)$
will, on this time scale, change non-continuously by a discrete value, $\Delta \dot{X}$. According to Eq$.$ (\ref{Trafo}), this change is given by
\begin{eqnarray}\label{Phasehopvelocity}
   \Delta\dot{X}(t_{\textrm{\scriptsize ph}})=\frac{1}{m\omega}V_1'(X(t_{\textrm{\scriptsize ph}}))\cdot\tilde{\Delta}\ ,
\end{eqnarray}
with
\begin{eqnarray}\label{Deltatilde} \tilde{\Delta}=\left[\int^{\tau}[f(\tau\!+\!\varphi\!+\!\Delta\varphi)]-\int^{\tau}[f(\tau\!+\!\varphi)]\right]_{\tau=\omega t_{\textrm{\tiny ph}}},
\end{eqnarray}
where $t_{\textrm{\scriptsize ph}}$ denotes the time at which the phase hop is induced. This change in the velocity of the particle's slow motion results in an instantaneous change $\Delta E$ (on the slow time scale) of the
period-averaged total energy $E$ of the particle, which is given by
\begin{eqnarray}\label{Phasehopenergy}
   \Delta
   E(t_{\textrm{\scriptsize ph}})=m\dot{X}(t_{\textrm{\scriptsize ph}})\Delta\dot{X}(t_{\textrm{\scriptsize
   ph}})+\frac{1}{2}m(\Delta \dot{X}(t_{\textrm{\scriptsize
   ph}}))^2,\
\end{eqnarray}
where $\dot{X}(t_{\textrm{\scriptsize ph}})$ denotes the velocity
of the slow motion right before the phase hop. As can be seen from
equations (\ref{Phasehopvelocity}) and (\ref{Phasehopenergy}),
$\Delta E(t_{\textrm{\scriptsize ph}})$ is of the same order
($\omega^{-2}$) as $E$. Thus, the period-averaged total energy
$E$ of a trapped particle can be significantly manipulated by a
phase hop for arbitrarily large $\omega$.\\
\indent  As an example, we
calculate $\Delta E$ for the special case
$f(\tau\!+\!\varphi)=\cos(\tau\!+\!\varphi)$, and assume that the
phase hop is induced at a time $t=t_{\textrm{\scriptsize ph}}$
for which $\int^{\tau}[\cos(\tau\!+\!\varphi)]_{\tau=\omega
   t_{\textrm{\tiny ph}}}=0$ (which is the case
twice each period of the fast motion), and for which the kinetic
energy of the slow motion equals three-quarters of the total period-averaged
energy, that is \mbox{$\frac{1}{2}m\dot{X}(t_{\textrm{\scriptsize
ph}})^2=\frac{3}{4}E$} (this is the case four times each period of the
slow motion). Due to energy conservation of the slow motion, it
is \mbox{$V_{\textrm{\scriptsize eff}}(X(t_{\textrm{\scriptsize
ph}}))\equiv \frac{1}{4m\omega^2}V_1'(X(t_{\textrm{\scriptsize
   ph}}))^2=\frac{1}{4}E$}. We further assume that the size of the
phase hop $\Delta\varphi$ equals \mbox{$\frac{\pi}{2}$}, such that
$\int^{\tau}[\cos(\tau\!+\!\varphi\!+\!\Delta\varphi)]_{\tau=\omega
   t_{\textrm{\tiny ph}}}=1$. Then
\begin{eqnarray}\label{deltaenergy}
   \Delta
   E(t_{\textrm{\scriptsize ph}})&=&m\dot{X}(t_{\textrm{\scriptsize ph}})\frac{1}{m\omega}V_1'(X(t_{\textrm{\scriptsize ph}}))+\frac{1}{2m\omega^2}V_1'(X(t_{\textrm{\scriptsize
   ph}}))^2\nonumber\\
   &=&\left(\frac{1}{2}\pm\sqrt{\frac{3}{2}}\right)E,
\end{eqnarray}
where the minus sign holds for $\dot{X}(t_{\textrm{\scriptsize ph}})\,V_1'(X(t_{\textrm{\scriptsize ph}}))<0$ and the plus sign for $\dot{X}(t_{\textrm{\scriptsize ph}})\,V_1'(X(t_{\textrm{\scriptsize ph}}))>0$. $\Delta E$ is therefore independent of $\omega$ and proportional to $E$, so that the relative change, \mbox{$\frac{\Delta E}{E}$}, is independent of $E$ (note, that this holds only in the limit of large $\omega$ and small $E$). It is therefore possible to take away more than two thirds of a trapped particle's energy or to give more than five thirds of its energy by inducing a phase hop for arbitrarily high frequencies and -- in the framework of classical physics -- for arbitrarily low energies (for numerical examples see Sec$.$ III, Fig$.$ \ref{Singlephasehop} and Fig$.$ \ref{Phasehopseries}). In Sec$.$ IV, we propose an experiment which may verify this result.

\section{Numerical Examples}
\begin{figure}[h]
\centering
\includegraphics[width=4.25cm]{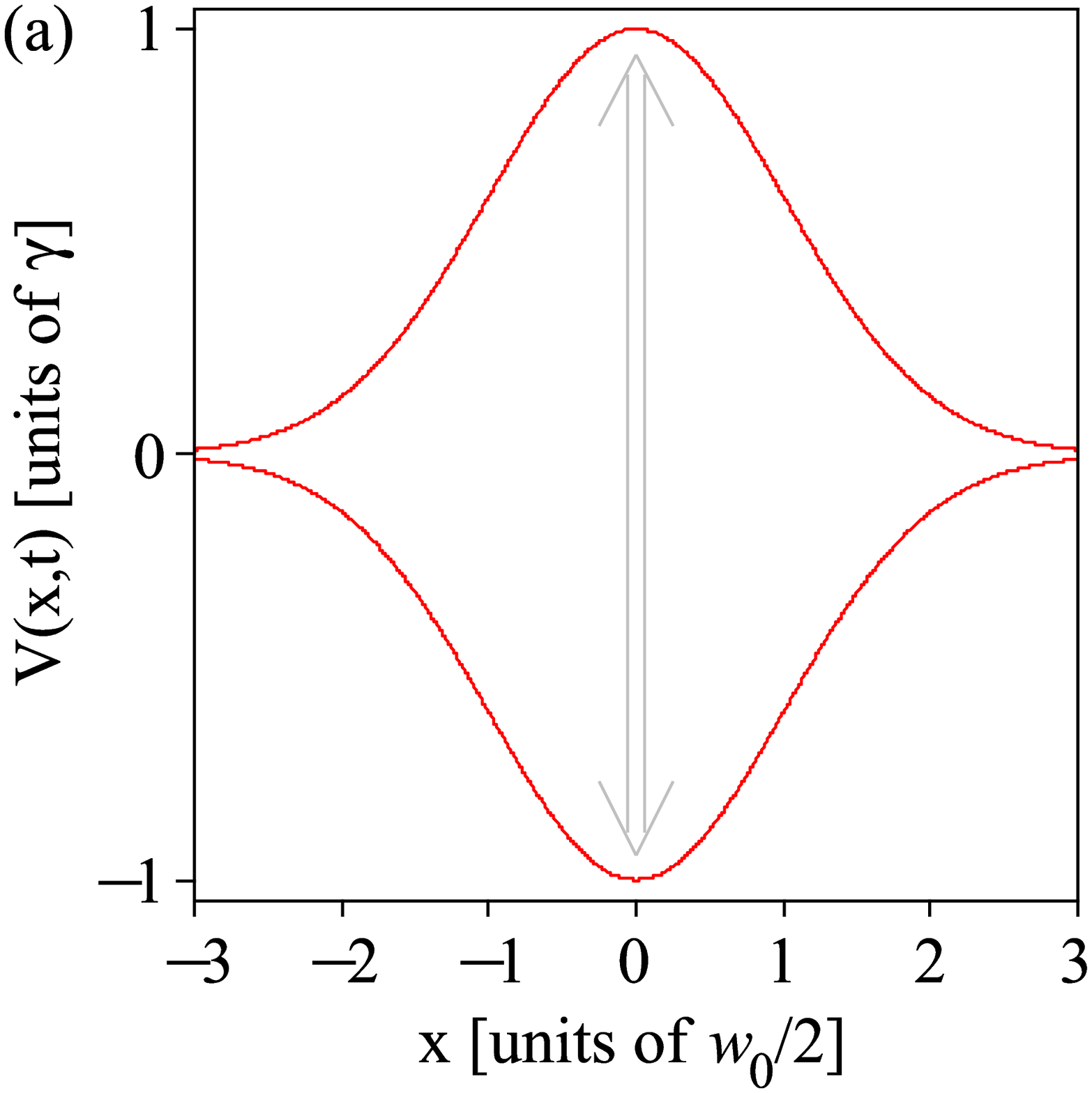}
\includegraphics[width=4.25cm]{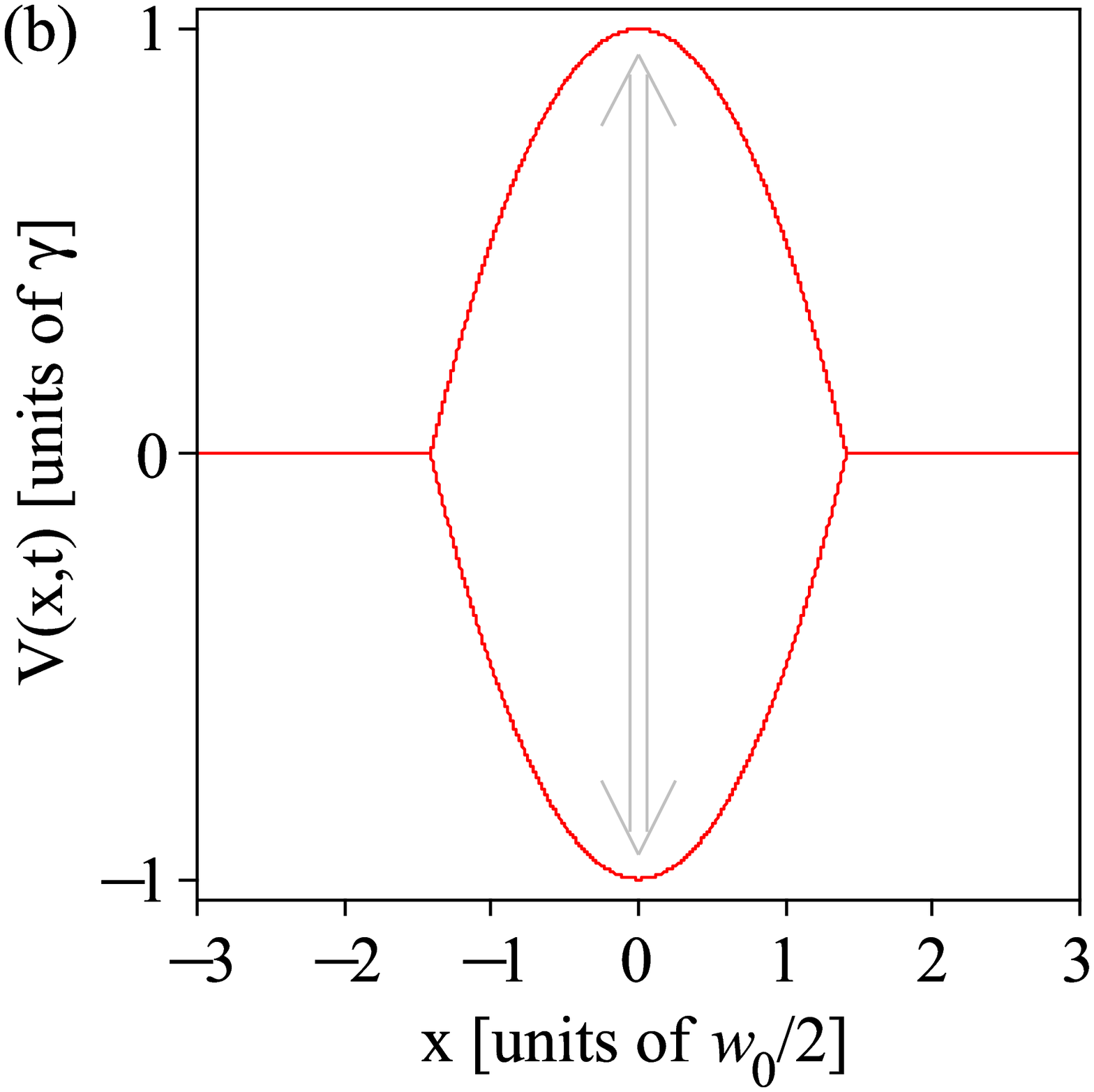}\\
\vspace{-0.3cm} \caption{(Color online) (a) Oscillating
Gaussian-shaped potential (\ref{Gaussianpotential}) and (b)
oscillating clipped parabola-shaped potential
(\ref{ParabolaPotential}).} \label{Demo}
\end{figure}

\begin{figure}[h]
\centering
\includegraphics[width=8.5cm]{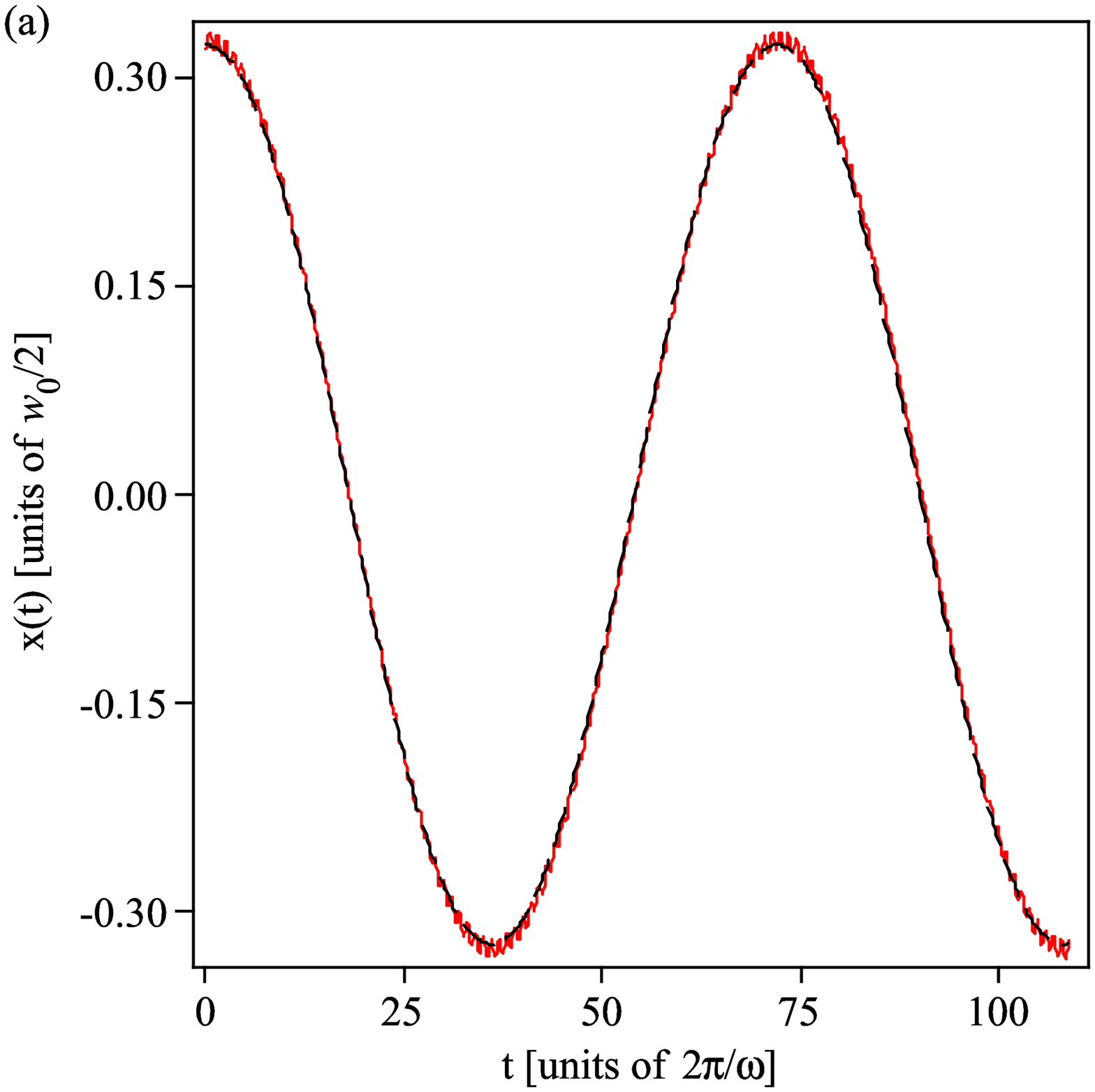}\\
\includegraphics[width=8.5cm]{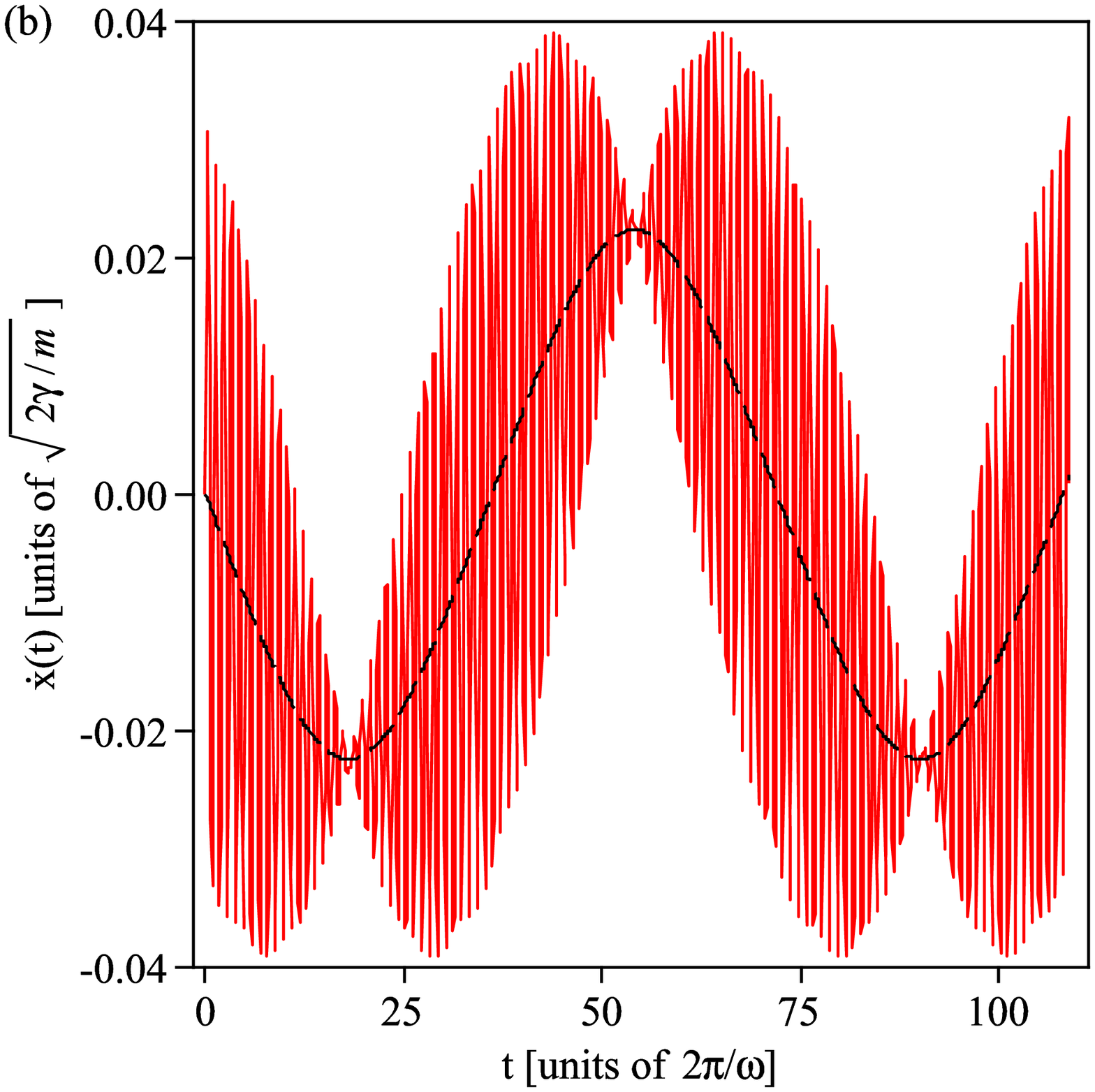}\\
\vspace{-0.3cm} \caption{(Color online) Trajectory and period-averaged trajectory of a particle
inside the oscillating Gaussian-shaped potential
(\ref{Gaussianpotential}) obtained from numerical simulations (a)
in position space and (b) in velocity space. Solid curve (red): real
trajectory, dashed curve (black): period-averaged trajectory. Parameters:
$\omega\!=\!7\,\omega_{\textrm{\scriptsize osc}}$; $\varphi=0$.
Initial conditions: $x(0)=0.32\,w_0$; $\dot{x}(0)=0$. The real and
period-averaged trajectory are almost identical in position space
but largely differ in velocity space.}\label{Trajectories}
\end{figure}

\begin{figure}[h]
\centering
\includegraphics[width=8.5cm]{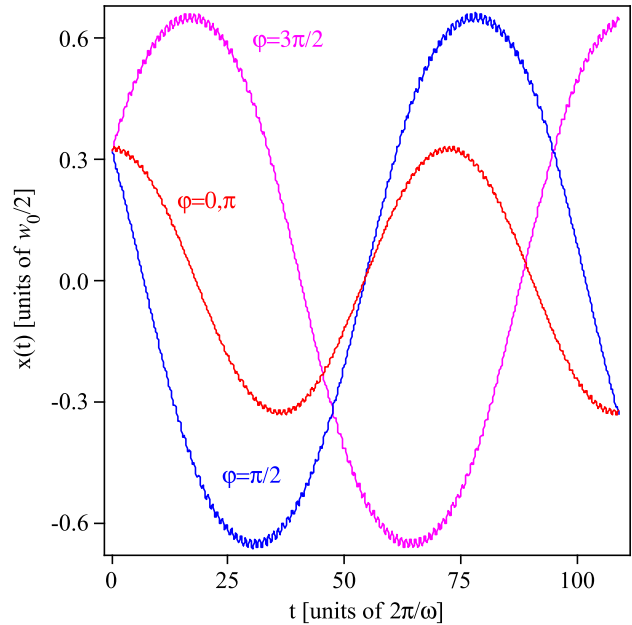}\\
\vspace{-0.3cm} \caption{(Color online) Trajectories in position space
of a particle inside the oscillating Gaussian-shaped potential
(\ref{Gaussianpotential}) for different $\varphi$, obtained from
numerical simulations. Parameters:
$\omega\!=\!7\,\omega_{\textrm{\scriptsize osc}}$ and
$\varphi\!=\!0$ (red curve), $\varphi\!=\!\mbox{$\frac{\pi}{2}$}$
(blue curve), $\varphi\!=\!\pi$ (also red curve), and $\varphi\!=\!\mbox{$\frac{3\pi}{2}$}$ (purple
curve). Initial conditions: $x(0)=0.32\,w_0$; $\dot{x}(0)=0$. The
shown trajectories are almost indistinguishable from their
theoretical predictions calculated from Eqs$.$ (\ref{Effectivepotential})
and (\ref{Trafo}), which are therefore not shown in the
figure.}\label{Trajectoriesdifferentphi}
\end{figure}

\begin{figure}[h]
\centering
\includegraphics[width=8.5cm]{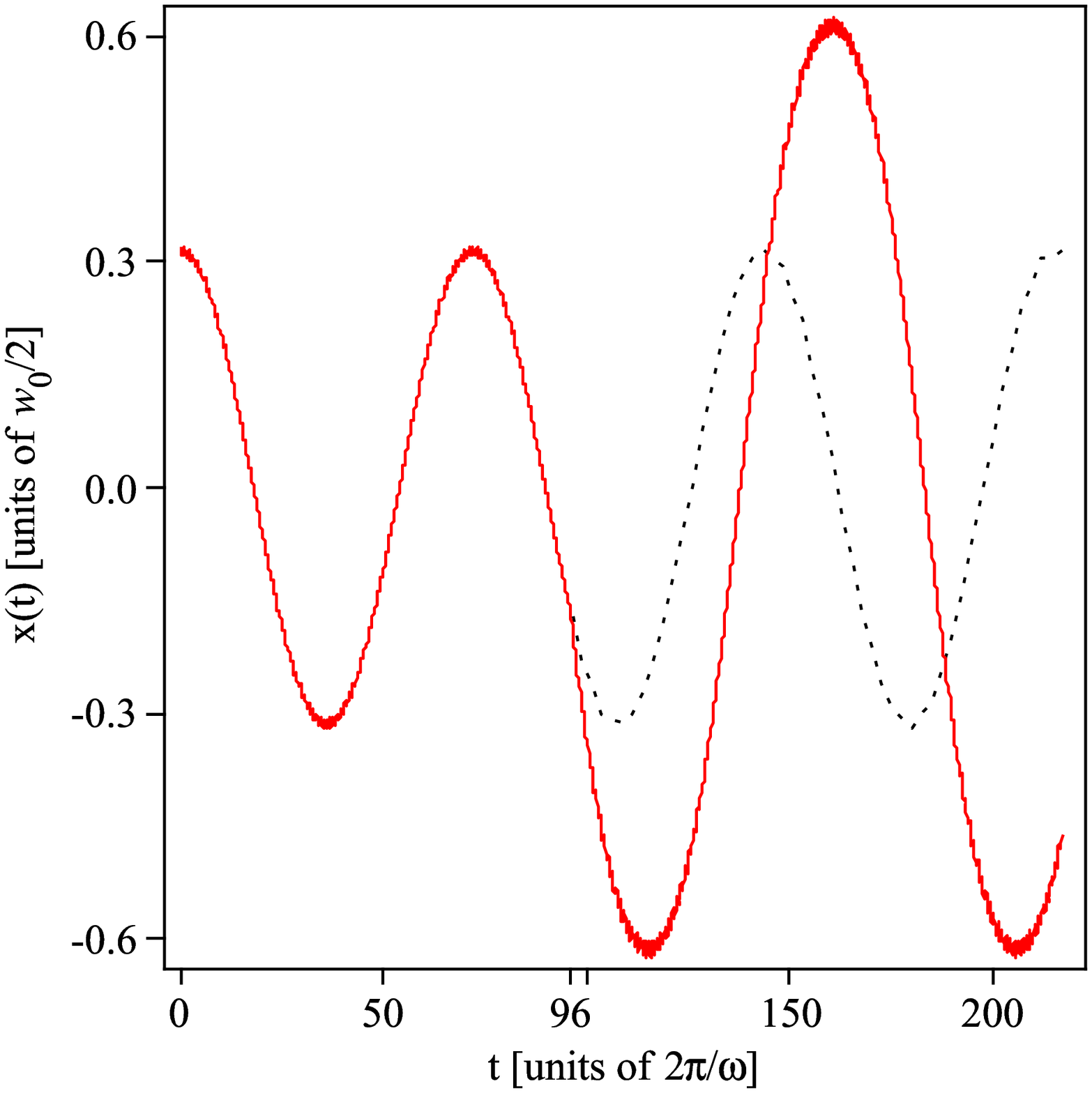}\\
\vspace{-0.3cm} \caption{(Color online) Trajectory in position space
of a particle inside the oscillating Gaussian-shaped potential
(\ref{Gaussianpotential}) when a phase hop is induced, obtained
from numerical simulations. Solid curve (red): real trajectory,
dotted curve (black): imaginary continuation of trajectory when
no phase hop was induced. Parameters:
$\omega\!=\!7\,\omega_{\textrm{\scriptsize osc}}$,
$\varphi\!=\!\pi$, time of phase hop $t_{\textrm{\scriptsize
ph}}\!=\!\mbox{$96\cdot\frac{2\pi}{\omega}$}$, and phase hop size
$\Delta\varphi\!=\!\mbox{$\frac{\pi}{2}$}$. Initial conditions:
$x(0)=0.32\,w_0$; $\dot{x}(0)=0$. The phase hop changes the
particle's total energy by a factor of approximately 2.7 in
agreement with equation (\ref{deltaenergy}). The shown trajectory
is almost indistinguishable from its theoretical prediction (not
shown) calculated from Eqs$.$ (\ref{Effectivepotential}) and
(\ref{Trafo}).}\label{Singlephasehop}
\end{figure}

\begin{figure}[h]
\centering
\includegraphics[width=8.5cm]{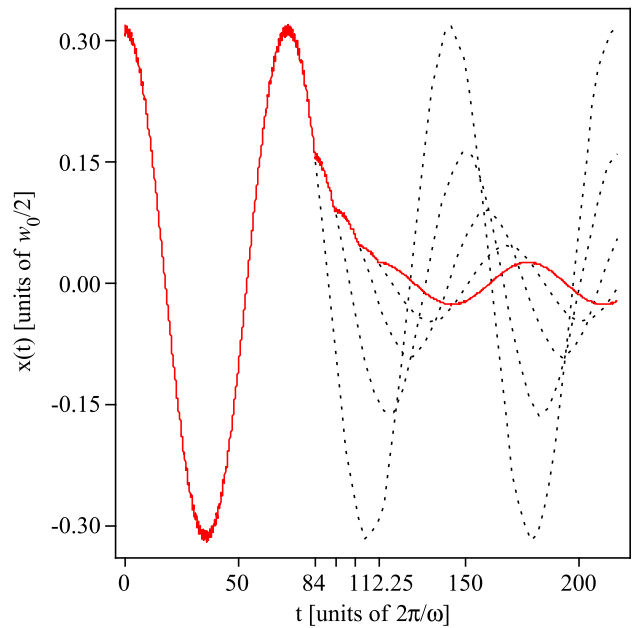}\\
\vspace{-0.3cm} \caption{(Color online) Trajectory in position space
of a particle inside the oscillating Gaussian-shaped potential
(\ref{Gaussianpotential}) when a series of four phase hops is
induced, obtained from numerical simulations. Solid curve (red):
real trajectory, dotted curves (black): imaginary continuations
of trajectory when the phase hops were not induced. Parameters:
$\omega\!=\!7\,\omega_{\textrm{\scriptsize osc}}$,
$\varphi\!=\!\pi$, times of phase hops
$t^{(1)}_{\textrm{\scriptsize
ph}}\!=\!\mbox{$84\cdot\frac{2\pi}{\omega}$}$,
$t^{(2)}_{\textrm{\scriptsize
ph}}\!=\!\mbox{$92.75\cdot\frac{2\pi}{\omega}$}$,
$t^{(3)}_{\textrm{\scriptsize
ph}}\!=\!\mbox{$103.5\cdot\frac{2\pi}{\omega}$}$,
$t^{(4)}_{\textrm{\scriptsize
ph}}\!=\!\mbox{$112.25\cdot\frac{2\pi}{\omega}$}$, and phase hop
size $\Delta\varphi\!=\!\mbox{$\frac{\pi}{2}$}$. Initial
conditions: $x(0)=0.32\,w_0$; $\dot{x}(0)=0$. The phase hops
change the particle's total energy by a factor of approximately
$1/170$ in agreement with equation (\ref{deltaenergy}). The shown
trajectory is almost indistinguishable from its theoretical
prediction (not shown) calculated from Eqs$.$ (\ref{Effectivepotential})
and (\ref{Trafo}).}\label{Phasehopseries}
\end{figure}

\begin{figure}[h]
\centering
\includegraphics[width=8.5cm]{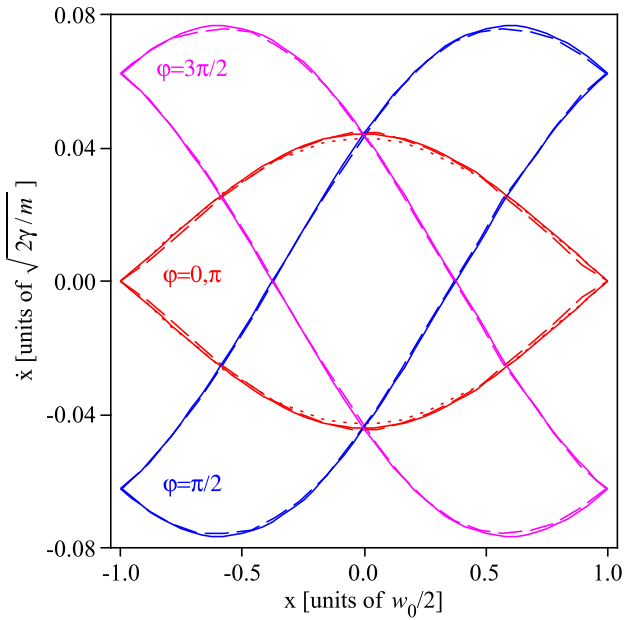}\\
\vspace{-0.3cm} \caption{(Color online) Stability regions for a
particle inside the oscillating Gaussian-shaped potential
(\ref{Gaussianpotential}) for different phases $\varphi$ and for
$\omega\!=\!7\,\omega_{\textrm{\scriptsize osc}}$. Solid curves:
Analytical predictions (\ref{Stabilityregionsgaussian}) for
$\varphi\!=\!0$ (red), $\varphi\!=\!\mbox{$\frac{\pi}{2}$}$
(blue), $\varphi\!=\!\pi$ (also red), and
$\varphi\!=\!\mbox{$\frac{3\pi}{2}$}$ (purple). Dashed curves:
Results of numerical simulations for $\varphi\!=\!0$ (red),
$\varphi\!=\!\mbox{$\frac{\pi}{2}$}$ (blue), and
$\varphi\!=\!\mbox{$\frac{3\pi}{2}$}$ (purple). Dotted curve:
For $\varphi\!=\!\pi$ (red).}\label{StabReg}
\end{figure}

\begin{figure}[h]
\centering
\includegraphics[width=8.5cm]{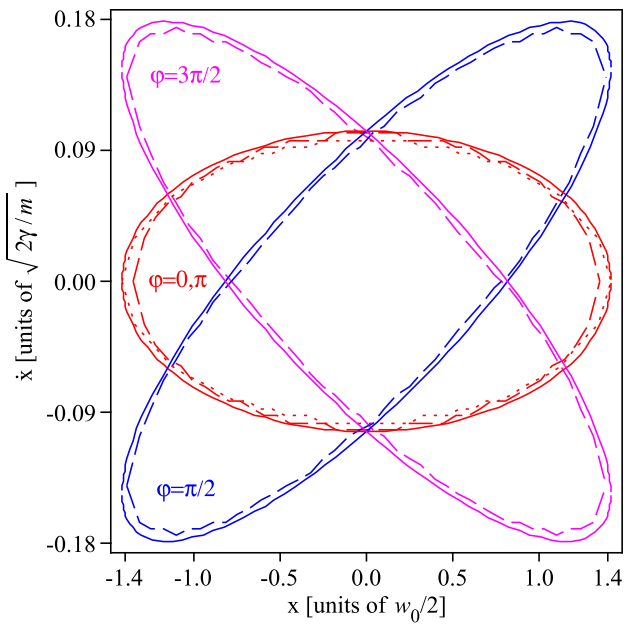}\\
\vspace{-0.3cm} \caption{(Color online) Stability regions for a
particle inside the oscillating clipped parabola-shaped potential
(\ref{ParabolaPotential}) for different phases $\varphi$ and for
$\omega\!=\!7\,\omega_{\textrm{\scriptsize osc}}$. Solid curves:
Analytical predictions
(\ref{StabilityRegionParabola}) for $\varphi\!=\!0$ (red),
$\varphi\!=\!\mbox{$\frac{\pi}{2}$}$ (blue), $\varphi\!=\!\pi$
(also red), and $\varphi\!=\!\mbox{$\frac{3\pi}{2}$}$ (purple).
Dashed curves: Results of Mathieu's theory for $\varphi\!=\!0$
(red), $\varphi\!=\!\mbox{$\frac{\pi}{2}$}$ (blue), and
$\varphi\!=\!\mbox{$\frac{3\pi}{2}$}$ (purple). Dotted curve:
For $\varphi\!=\!\pi$ (red).}\label{StabReg2}
\end{figure}

In this section, we consider two simple examples in order to
demonstrate the phenomenon found in the preceding section and to
compare the derived analytical predictions to results of numerical
simulations. The first example to be considered is a particle
moving inside an oscillating Gaussian-shaped potential with a
vanishing time average of the form
\begin{eqnarray}\label{Gaussianpotential}
    V^{\textrm{\scriptsize G}}(x,t)=\gamma\exp\left(-\frac{2x^2}{w_0^2}\right)\cos(\omega t+\varphi),
\end{eqnarray}
which is illustrated in Fig$.$ \ref{Demo}(a).
%
%
Potential (\ref{Gaussianpotential}) defines two time scales: The time scale associated with the
frequency $\omega$ of the potential's oscillation and the time scale associated with the oscillation frequency $\omega_{\textrm{\scriptsize
osc}}$ of a low energy particle in the minimum instantaneous potential. The theory which was derived in Sec$.$ II is valid for large frequencies $\omega$ compared to the frequencies which are associated with the other time scales of the system. In the following, we therefore consider the regime $\omega\!\gg\!\omega_{\textrm{\scriptsize osc}}$.\\
\indent In the preceding section, we showed theoretically, that for correctly chosen initial conditions the position of a trapped particle can be approximated by the position of its slow motion, whereas its velocity cannot be approximated by the velocity of its slow motion. To verify this, we numerically simulate a particle's trajectory in position and velocity space and compare it to its period-average. This is done by numerically integrating the time dependent Newton equation
\begin{eqnarray}\label{NewtonequationGaussian}
   m\ddot{x}=\frac{4\gamma}{w_0^2}x\exp\left(-\frac{2x^2}{w_0^2}\right)\cos(\omega t+\varphi).
\end{eqnarray}
The results are shown in Fig$.$ \ref{Trajectories}.
%
%
In the figure, one can clearly see that the trajectory in position space is well described by its period-average (Fig$.$ \ref{Trajectories}(a)), whereas in velocity space (Fig$.$ \ref{Trajectories}(b)) this is not the case. Additional numerical simulations showed that this result does not change when $\omega$ is increased, in agreement with the derived theory.\\
\indent In Sec$.$ II, we showed that the significant difference between the particle's velocity and its period-average (Fig$.$ \ref{Trajectories}(b)) leads to a dependence of the particle's motion on the initial phase $\varphi$ of the oscillating potential. To verify this conclusion, we numerically simulate the particle's trajectory in position space for different $\varphi$ for a given frequency $\omega$ and given initial conditions of the particle. The result is shown in Fig$.$ \ref{Trajectoriesdifferentphi}.
%
%
The charted trajectories appreciably differ from each other, as predicted by the derived theory. On the slow time scale they have different phases and different amplitudes. Figure \ref{Trajectoriesdifferentphi} shows as well, that all trajectories appear to be generated from the same time independent effective potential. Since the trajectories for different $\varphi$ are not the same although their initial conditions are, this verifies that the trajectories are determined in addition by a transformation of their initial conditions which depends on $\varphi$. The shown trajectories are almost indistinguishable from their theoretical predictions calculated from equations (\ref{Effectivepotential})
and (\ref{Trafo}), which are therefore not shown in the figure.\\
\indent Another result of Sec$.$ II to be verified by numerical simulations is the prediction of the particle's motion when phase hops in the oscillating potential's time modulation function are induced. Figure \ref{Singlephasehop} shows the numerically simulated trajectory of the particle for a phase hop of size $\Delta\varphi\!=\!\mbox{$\frac{\pi}{2}$}$ induced at a time $t=t_{\textrm{\scriptsize ph}}$ for which $\int^{\tau}[\cos(\tau\!+\!\varphi)]_{\tau=\omega
   t_{\textrm{\tiny ph}}}=0$, \mbox{$\frac{1}{2}m\dot{X}(t_{\textrm{\scriptsize
ph}})^2\approx\frac{3}{4}E$}, and $\dot{X}(t_{\textrm{\scriptsize ph}})\,V_1'(X(t_{\textrm{\scriptsize ph}}))>0$.
%
%
In agreement with the analytical prediction (\ref{deltaenergy}), the
period-averaged energy of the particle in Fig$.$
\ref{Singlephasehop} indeed changes by a factor of approximately
$2.7$. Furthermore, the shown trajectory is almost indistinguishable from its theoretical prediction (not shown in the figure) calculated from equations (\ref{Effectivepotential}) and (\ref{Trafo}), whereas the part of the trajectory after the
phase hop is obtained from these by assigning the time of
the phase hop to the initial time.\\
\indent Figure \ref{Phasehopseries} shows the numerically simulated trajectory of
the particle when several phase hops of the size $\Delta\varphi\!=\!\mbox{$\frac{\pi}{2}$}$ are induced. The moments of the phase hops $t_i=t^{(i)}_{\textrm{\scriptsize ph}}$ were chosen such that 
$\int^{\tau}[\cos(\tau\!+\!\varphi)]_{\tau=\omega
   t^{(i)}_{\textrm{\tiny ph}}}=0$, \mbox{$\frac{1}{2}m\dot{X}(t^{(i)}_{\textrm{\scriptsize
ph}})^2\approx\frac{3}{4}E(t^{(i)}_{\textrm{\scriptsize
ph}})$}, and $\dot{X}(t^{(i)}_{\textrm{\scriptsize ph}})\,V_1'(X(t^{(i)}_{\textrm{\scriptsize ph}}))<0$, whereas $X(t^{(i)}_{\textrm{\scriptsize ph}})$, $\dot{X}(t^{(i)}_{\textrm{\scriptsize ph}})$, and $E(t^{(i)}_{\textrm{\scriptsize
ph}})$ denote the position, the velocity, and the energy of the particle's slow motion right before the $i$th phase hop, respectively.
%
%
In agreement with the analytical prediction (\ref{deltaenergy}), the period-averaged energy of the particle in Fig$.$ \ref{Phasehopseries} changes in total by a factor of approximately \mbox{$\frac{1}{170}$}. Figure \ref{Phasehopseries} demonstrates that this substantial change in energy can be achieved even when the phase hops are induced very quickly one after another (within a fraction of the period of the particle's slow motion). The shown trajectory is almost indistinguishable from its theoretical prediction (not shown) calculated from Eqs$.$ (\ref{Effectivepotential}) and (\ref{Trafo}).\\
\indent The last theoretical prediction of Sec$.$ II that is to be verified by numerical simulations in the following is the derived equation (\ref{Thesis}) for the stability region. For the particle inside the oscillating Gaussian-shaped potential (\ref{Gaussianpotential}), it yields
\begin{eqnarray}\label{Stabilityregionsgaussian}
\!\!\!\!\!\!\!\!\!\!\dot{x}^{\textrm{\scriptsize (G)}}\!\!&\lessgtr&\!\!
-\frac{1}{m\omega}\frac{4\gamma}{w_0^2}x\exp\left(\mbox{$-\frac{2x^2}{w_0^2}$}\right)\sin\varphi\nonumber\\
&&\!\!\pm\frac{1}{m\omega}\frac{\sqrt{8}\gamma}{w_0^2}\sqrt{\frac{w_0^2}{4}\exp\left(\mbox{$-1$}\right)-x^2\exp\left(\mbox{$-\frac{4x^2}{w_0^2}$}\right)}.
\end{eqnarray}
A graph of this analytical prediction of the stability region for
different $\varphi$ is shown in Fig$.$ \ref{StabReg} (solid
curves). Before comparing it to the
results of numerical simulations, we first discuss it separately.
%
%
In Fig$.$ \ref{StabReg}, one can clearly see that the analytically
predicted stability regions for different $\varphi$ significantly
differ from each other. For the initial phases $\varphi\!=\!0$ and
$\varphi\!=\!\pi$, the stability regions are identical and axially
symmetrical to the coordinate axes. For the phases
$\varphi\!=\!\mbox{$\frac{\pi}{2}$}$ and
$\varphi\!=\!\mbox{$\frac{3\pi}{2}$}$, the stability regions are
not axially symmetrical to the coordinate axes and particularly
not to the position axis. This demonstrates, that the motion of
the particle cannot be determined by a time independent effective
potential alone, since the stability region of any time
independent potential is due to energy conservation axially
symmetric to the position axis. Compared to the stability regions
of the phases $\varphi\!=\!0$ and $\varphi\!=\!\pi$, those of the
phases $\varphi\!=\!\mbox{$\frac{\pi}{2}$}$ and
$\varphi\!=\!\mbox{$\frac{3\pi}{2}$}$ further look distorted and
rotated around the origin. One can see, that the area of the
stability region is independent of the initial phase $\varphi$,
which immediately emerges from equation (\ref{Thesis}). The overlap
of two stability regions of different $\varphi$ in units of their
total area, however, depends on the initial phase of the
respective overlapping stability regions, but is according to Eq$.$
(\ref{Thesis}) independent of $\omega$. The spatial width of the
stability region is independent of the initial phase $\varphi$ and
according to Eq$.$ (\ref{Thesis}) also independent of $\omega$. Finally, Fig$.$ \ref{StabReg} shows that the boundaries of all stability regions intersect at $x\!=\!0$, and thus the escape velocity
of the particle that is initially situated at $x\!=\!0$ is independent of the initial phase $\varphi$, in agreement with the numerical results presented in \cite{AbrCha04}. This is, because for $x(0)\!=\!0$ the initial condition transformation (\ref{Trafo}) is
independent of $\varphi$, since it is
$V_1'(0)\!=\!0$.\\
\indent We now compare the analytical prediction for the stability region (\ref{Stabilityregionsgaussian}) to results of numerical simulations, which are shown as dashed and dotted curves in Fig$.$ \ref{StabReg}. The numerical
simulations were performed as follows: First, the particle's
phase space was discretized into a grid of equally spaced points,
each of which define possible initial conditions of the
particle. Then, the time dependent Newton equation (\ref{NewtonequationGaussian})
was integrated numerically for each of these points. Thus, all the
points, which generated a trajectory that stayed within a
prespecified spatial region within a prespecified \textit{large}
time, were defined as points of the stability region. The
prespecified values for the spatial region
($|x|\!\lesssim\!\mbox{$\frac{w_0}{2}$}$) and the \textit{large} time ($t\!\approx\!\mbox{$200\cdot\frac{2\pi}{\omega}$}$) were
chosen such that further increase in these values did not change
the result appreciably. The step size was chosen to be $\Delta
t\!=\!\mbox{$\frac{1}{100}\!\cdot\!\frac{2\pi}{\omega}$}$, since
further decrease of the step size did not alter the results
appreciably. Figure \ref{StabReg} shows a very good agreement between theory and
simulations. Additional simulations showed, that this agreement
increases when $\omega$ is increased, as expected from the theory.\\
\indent So far we have considered a particle moving inside the
oscillating Gaussian-shaped potential (\ref{Gaussianpotential}).
Another example to be considered in the following is a particle
moving inside an oscillating clipped parabola-shaped potential
with a vanishing time average of the form
\begin{eqnarray}\label{ParabolaPotential}
    V^{\textrm{\scriptsize cp}}(x,t)=V_1(x)\cos(\omega t+\varphi),
\end{eqnarray}
with
\begin{eqnarray}\label{ParabolaPotentialmoreexplicit}
    V_1(x)=\begin{cases}-kx^2+kL^2 & \mbox{for $|x|\leq L$,}\\ 0 &
    \mbox{else,}\end{cases}
\end{eqnarray}
which is illustrated in Fig$.$ \ref{Demo}(b). This is a problem
able to be solved exactly, because the particle's time dependent
Newton equation
\begin{eqnarray}\label{NewtonequationParabola}
   m\ddot{x}=\begin{cases}2kx\cos(\omega t+\varphi) & \mbox{for $|x|\leq L$,}\\ 0 &
    \mbox{else}\end{cases}
\end{eqnarray}
is for $|x|\!\leq\!L$ equivalent to the homogeneous Mathieu
equation, whose solution is given by a linear combination of the
Mathieu functions \cite{Pau90,LeiBla03,MajGhe04}, and for
$|x|\!\geq\!L$ it is equivalent to the trivial differential
equation $\ddot{x}=0$ of free motion. Therefore, the considered
example allows for a comparison between the theory that was
derived in Sec$.$ II and the Mathieu theory. We choose
$L=\mbox{$\frac{w_0}{\sqrt{2}}$}$ and
$k=\mbox{$\frac{\gamma}{L^2}$}$ so that in
Eq$.$ (\ref{ParabolaPotential}) the reference frequency
$\omega_{\textrm{\scriptsize
osc}}\!=\!\sqrt{\frac{4\gamma}{mw_0^2}}$ as well as the maximum
instantaneous potential depth $\gamma$ are the same as in
(\ref{Gaussianpotential}). In the following, we calculate the
particle's stability region -- once by using the derived equation (\ref{Thesis}) and once by applying Mathieu's theory.
Equation (\ref{Thesis}) yields
\begin{eqnarray}\label{StabilityRegionParabola}
\!\!\dot{x}^{\textrm{\scriptsize (cp)}}\lessgtr
-\frac{1}{m\omega}\frac{4\gamma}{w_0^2}x\sin(\varphi)\pm\frac{1}{m\omega}\frac{2\sqrt{2}\gamma}{w_0^2}\sqrt{\frac{w_0^2}{2}-x^2}.
\end{eqnarray}
A graph of this analytical prediction of the stability region is
shown in Fig$.$ \ref{StabReg2} (solid curves) for different
$\varphi$. Figure \ref{StabReg2} also shows the results obtained
from Mathieu's theory (dashed and dotted curves).
%
%
It can be seen that the stability regions of the particle have the
shape of an ellipse. Figure \ref{StabReg2} shows a very good
agreement of our derived theory and the Mathieu theory. In order
to confirm the accuracy of the numerical simulations which we
performed for the oscillating Gaussian-shaped potential in the
preceding paragraph, we repeated these simulations for the
parabola-shaped potential finding that the obtained results were
identical to those obtained from Mathieu's theory. All main
results which we demonstrated for the oscillating Gaussian-shaped
potential were found also for the oscillating parabola-shaped
potential.

\section{Proposal of an experiment}
\begin{figure}[h]
\centering
\includegraphics[width=8.5cm]{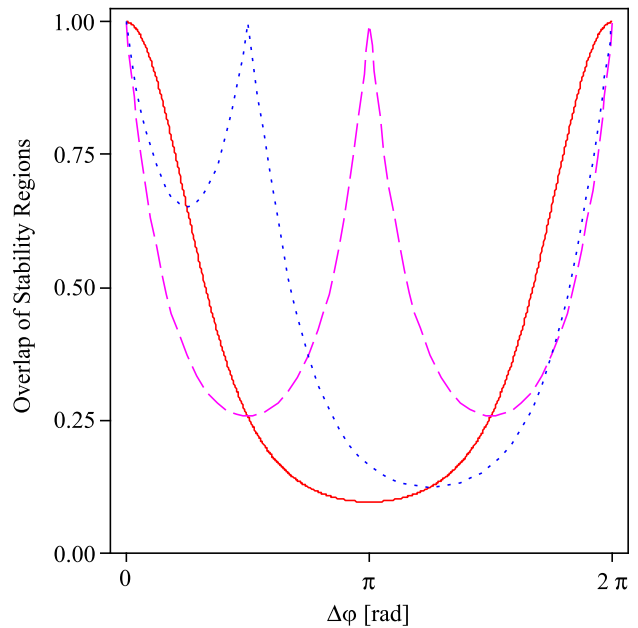}\\
\vspace{-0.3cm} \caption{(Color online) Overlap of the stability
region of initial phase $\varphi$ with the stability region of
initial phase $\varphi\!+\!\Delta\varphi$ for different
$t_{\textrm{\scriptsize ph}}$ as a function of $\Delta\varphi$
obtained from the two-dimensional extension of equation
(\ref{Thesis}) for an oscillating Gaussian-shaped potential. Solid curve (red):
$(\omega t_{\textrm{\scriptsize ph}}\!+\!\varphi) \textrm{
mod }2\pi\!=\!\mbox{$\frac{\pi}{2}$}$, dashed curve (purple): $(\omega
t_{\textrm{\scriptsize ph}}\!+\!\varphi) \textrm{ mod
}2\pi\!=\!\pi$, dotted curve (blue): $(\omega
t_{\textrm{\scriptsize ph}}\!+\!\varphi) \textrm{ mod
}2\pi\!=\!\mbox{$\frac{3\pi}{4}$}$. The overlap of the stability regions is given in
units of their total volume.} \label{Overlap}\end{figure}

In this section, we propose an experiment to verify our
theoretical results. Consider an ensemble of atoms that is trapped
inside a rapidly oscillating potential with a vanishing time
average, whose amplitude is chosen such that the ensemble
uniformly fills its stability region. Hence the fraction of atoms
which remain trapped after a phase hop of the oscillating
potential is induced is equal to the mutual overlap of the
stability region with the initial time given by the time right
before the phase hop and the stability region with the initial
time given by the time right after the phase hop. This overlap is
exactly determined by equation (\ref{Thesis}). It is a function of
the time $t_{\textrm{\scriptsize ph}}$ when the phase hop is
induced and of the size $\Delta \varphi$ of the phase hop. Figure
\ref{Overlap} shows this overlap for a two-dimensional oscillating
Gaussian-shaped potential.
%
%
One can see that, regardless of $t_{\textrm{\scriptsize ph}}$, at least $70\%$ of the atoms will leave the trap for appropriate
$\Delta \varphi$. Therefore, when measuring the ratio between the numbers of trapped atoms before and after the phase hop for different phase hop sizes, one can expect a clear signal from the experiment.

\section{Conclusion}
In conclusion, we investigated the classical dynamics of a particle
moving inside a rapidly oscillating potential with a vanishing
average. We found, that the motion of such a particle
significantly depends on the oscillating potential's initial phase
in the case that the particle is trapped. The effective
potential theory described in the literature failed to describe this phenomenon. We found that this theory is incomplete. It states that the motion of the particle is determined by a time independent effective potential, whereas we could show that this is true only if the particle's initial conditions are transformed according to a non-negligible transformation which is significant even for arbitrarily high frequencies of the potential's oscillation. We derived this transformation and showed that it depends on the potential's initial phase which explains the found phenomenon. We also showed that there always exist two initial phases of the oscillating potential such that the transformation can be
neglected. Therefore, the effective potential theory in its
previous form, which omits this transformation, can only describe
these two special cases correctly. The phase dependence offers a
new possibility to significantly manipulate the dynamics of a
trapped particle. We explicitly calculated the change in a
particle's energy caused by a phase hop in the
potential's modulation function. We found that it is significant for arbitrarily high frequencies of the potential's oscillation and -- in the framework of classical physics -- for arbitrarily low energies of the particle. It is the subject of future research
to find applications of this novel tool of particle manipulation.
The results presented in this article, should also be extended to
the quantum regime.

\section*{ACKNOWLEDGMENTS}
We would like to thank Shmuel Fishman, Roee Ozeri, Nir Bar-Gill, Rami Pugatch and Yoav Sagi for stimulating and inspiring discussions. This research was supported by the ``Stiftung der Deutschen Wirtschaft'' of the Federal Ministry of Education and Research (BMBF) of Germany and by the MINERVA Foundation.

\end{document}